\documentclass[reprint, superscriptaddress, aps, floatfix]{revtex4-2}
\usepackage[utf8]{inputenc}
\usepackage{xr}
\usepackage{hyperref}
\hypersetup{colorlinks=true, linkcolor=blue, filecolor=blue, urlcolor=blue, citecolor=blue}
\usepackage{amsmath}
\usepackage{cleveref}
\usepackage{multirow}
\usepackage{graphicx}
\usepackage{dcolumn}
\usepackage{bm}
\usepackage{xcolor}
\usepackage{textcomp}
\usepackage{pgf}
\usepackage{bm}

\usepackage{xcolor}

\DeclareUnicodeCharacter{2212}{\ensuremath{-}}
\DeclareUnicodeCharacter{2009}{\,}

\begin{document}

%\title{Multi-stage relaxation in model glass revealed by computational X-ray photon correlation spectroscopy}
%\title{Multi-stage structural decorrelation in a model glass}
\title{Microstructure-specific mechanisms define multistage relaxation dynamics in a metallic model-glass}

\author{Achraf Atila}
\affiliation{Federal Institute of Materials Research and Testing (BAM), Unter den Eichen 87, Berlin 12205, Germany}

\author{Zengquan Wang}
\affiliation{Federal Institute of Materials Research and Testing (BAM), Unter den Eichen 87, Berlin 12205, Germany}

\author{Birte Riechers}
\affiliation{Federal Institute of Materials Research and Testing (BAM), Unter den Eichen 87, Berlin 12205, Germany}

\author{Robert Maa{\ss}}
\email{robert.maass@bam.de}

\affiliation{Federal Institute of Materials Research and Testing (BAM), Unter den Eichen 87, Berlin 12205, Germany}

\affiliation{Department of Materials Engineering, Technical University of Munich, 85748 Garching, Germany}

\affiliation{Department of Materials Science and Engineering, University of Illinois at Urbana-Champaign, 61801 Urbana, USA}

\date{Submitted on 24 July 2025}

\begin{abstract}
Deciphering complex relaxation pathways in disordered solids is a central challenge across polymeric, oxide, and metallic glasses, which traditionally relies on the interpretation of mechanical spectroscopy and resulting damping modes. Here we demonstrate the direct observation of dominant atomic-scale relaxation mechanisms during isothermal annealing of an as-quenched binary model glass towards incipient crystallization. Assessed via simulated x-ray photon correlation spectroscopy, a multi-state structural decorrelation is uncovered via speckle-pattern analysis of the full three-dimensional diffraction sphere across the first peak of the structure factor. Over a simulation time of up to~10 $\mu$s, three distinct and subsequent decorrelation stages of thermal vibration, glassy network evolution, and structural and chemical ordering towards crystallization are identified. These findings promote a picture where specific dynamically-separated mechanisms drive the microstructural evolution during glass relaxation and suggest a much richer multi-mode relaxation behavior of metallic glasses than hitherto identified. 
\end{abstract}
%\linenumbers

\maketitle

Disordered solids, such as polymer glasses, oxide glasses, or metallic glasses, are out-of-equilibrium structures that relax with time. This is due to the inherent driving force of local atomic-scale energy minimization, where the asymptotic reference may be embodied in the negative excess-entropy state of the Kauzmann paradox~\cite{Kauzmann1948ChemRev}, a crystalline, or an at least partially crystalline solid~\cite{Zanotto2017JNCS}. Depending on the temperature and the type of glassy solid, the time-scales for structural relaxation range over many orders of magnitude. This, together with the difficulty in establishing general, yet well-defined structural mechanisms that underlie physical aging, makes glass dynamics and relaxation a forefront problem in condensed matter physics and materials science.

Initially quantified via simple viscoelastic Maxwell relaxation times~\cite{Doss2020JACS}, it is now known that glassy solids exhibit a rich spectrum of dominant relaxation modes, typically separated into $\alpha$-relaxation near the glass-transition temperature, $T_\text{g}$, and secondary modes at lower homologous temperatures~\cite{Huang2021NatCom, Chang2022NatMat}. The latter include the Johari-Goldstein $\beta$-relaxation~\cite{Goldstein1969}, the $\gamma$-mode~\cite{Kchemann2017Scripta}, and the $\delta$-mode~\cite{Boulougouris2009JCP}, some of which are generically observed in glassy solids~\cite{Goldstein1969, Schneider2000PRL, Hecksher2017PNAS, Kchemann2017Scripta, Chang2022NatMat, Gao2025NatP}, whereas others are limited to specific structural or chemical realizations. In fact, rather complex, multi-mode mechanical relaxation spectra in both polymer and oxide glasses are common~\cite{Guiselin2022NatP}, sharing the complexity also present in damping-spectra of defect-dominated crystalline systems~\cite{nowick2012anelastic}.

In the more modern class of metallic glasses, relaxation dynamics was for a long time limited to $\alpha$- and $\beta$-modes~\cite{Wang2019PMS}, and subsequently expanded to $\gamma$-relaxations~\cite{Kchemann2017Scripta}. Novel mechanical spectroscopy methods with significantly enlarged frequency ranges have very recently reported additional low frequency loss signatures, referred to $\alpha_2$~\cite{Sun2019PRL}, $\alpha$'~\cite{Luo2017PRL}, or $u$-relaxations~\cite{Yang2024APR}. These developments motivate the hypothesis of a spatially heterogeneous structure that enables a variety of structure-specific dominant relaxation pathways. Together, they build the overall broad barrier-energy spectrum of the disordered metallic solid~\cite{Rodney2009PRB, Fan2017NatCom}. 

Indeed, experiments provide indirect evidence for a complex and heterogeneous metallic glass (micro)structure beyond short-range order (SRO) length scales via numerous signatures of spatially-varying property~\cite{Riechers2023, Rashidi2025ScripMat} and relaxation-time fluctuations~\cite{Riechers2024}, enthalpy recovery~\cite{Hodge1994JNCS,Costa2023}, negative creep~\cite{Castellero2008,Derlet2021ActaMat}, and atomic-scale dynamics~\cite{Wang2024Acta}. However, in contrast to oxide and polymer glasses, direct insights into the structural details (atomic-motif network, bond structures, dissipation mechanisms, etc.) remain experimentally inaccessible, and only recently model-glass simulations at unprecedented microsecond-long simulation times have brought to light how microstructures may emerge due to thermally-activated transport~\cite{Derlet2018ActaMat, Derlet2021PRM}. Such a glass microstructure consists of low-energy motifs, for example, full icosahedra, and Frank-Kasper polyhedra that form extended clusters, where more frustrated domains are interspersed and admit transport via string excitations. The latter follow well-defined bond-order topologies~\cite{Derlet2020PRM}, admit Eshelby stress-fields~\cite{Derlet2021PRM}, and have a strong signature of correlated-collective dynamics in the time-domain~\cite{Riechers2024}. These characteristics underline the above proposition that defined structural environments develop during continued metallic glass relaxation, bearing the promise to map relaxation dynamics and their dominant processes onto unique structural components. 

The central challenge in proving this hypothesis is the need for an atomic-scale probe with sufficient spatial and dynamical sensitivity. Here the advent of either x-ray or electron-correlation methods have partly addressed the experimental needs~\cite{Shpyrko2014JSR, He2015MM}, providing time-resolved insights into the atomic-scale dynamics via the time-dependent quantification of the intermediate scattering function (ISF). Notably, pronounced heterogeneous structural dynamics in response to temperature~\cite{Ruta2012PRL}, stress~\cite{Das2019NatCom}, and as a function of position~\cite{Zhang2018NatCom} has been demonstrated, which is different from the volume-averaged smooth relaxation response obtained via classical stress-relaxation.

Despite this advance in understanding the complex coupling between metallic-glass structure and relaxation-time spectra, all available experimental data convey a picture of a single time-dependent decay of the ISF, typically fitted with a Kohlrausch-Williams-Watts (KWW) function and motivated by the Siegert approximation~\cite{Ferreira2020AJP}. This form implies a broad, continuous distribution of relaxation times, though an effective average timescale can still be defined over the chosen interrogation-time and resolution. Given the above hypothesis, a true glassy microstructure would indeed suggest a quite different scenario where different microstructural domains admit sufficiently well-separated dynamic signatures.

\begin{figure*}
\centering
\includegraphics[width=0.8\textwidth]{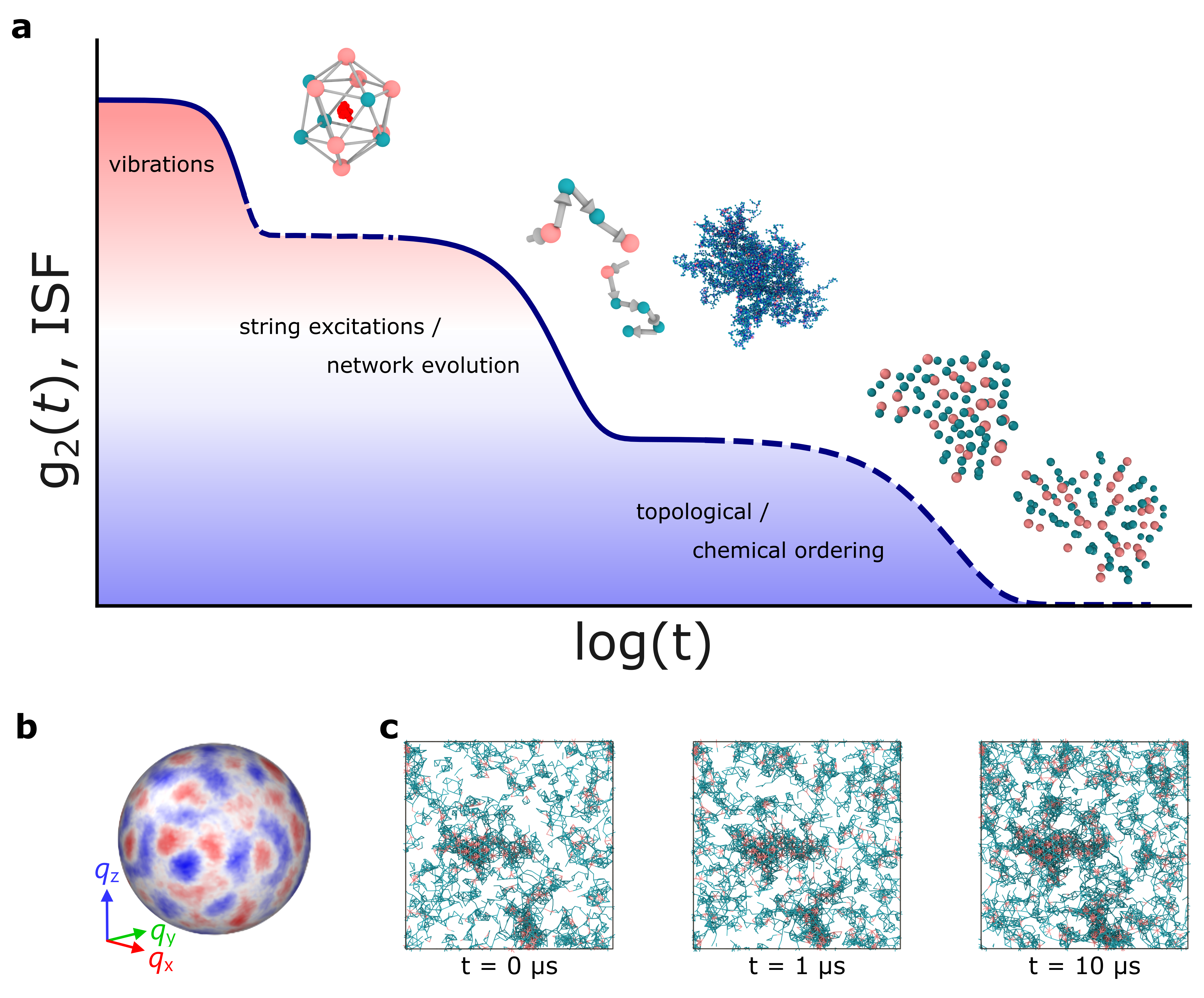}
\caption{(a) Schematic of the here uncovered multistage decorrelation of the studied model glass. Visualizations of the dominant mechanism that drive structural decorrelation at different time-scales are indicated: at short timescales, the decorrelation is governed by thermal vibrations (depicted by the trajectory of the central atom of a selected icosahedron), while at intermediate times a second decay emerges through collective string-like motion and the associated  evolution of a network, which is a signature of $\beta$-relaxation (shown through two selected strings and an evolved large cluster). The last relaxation domain is a consequence of structural fragments that order topologically and chemically, and that indicate incipient crystallization. (b) Visualization the three-dimensional structure factor obtained from MD simulations of the model binary glass for $7 < q < 8$ $\sigma^{-1}$. (c) Snapshots of slices of thickness 5 $\sigma$ that show the evolution of the low-energy (IFK) motif network in the simulation box taken at times $t = $0, 1, and 10 $\mu$s at 0.345~$k_\text{B}T$. These snapshots only show atoms belonging to low-energy (IFK) motif network, while atoms of the more frustrated amorphous structure have been removed for visualization purposes.}
\label{fig:Figure1}
\end{figure*}

To this end, we pursue here simulated x-ray photon correlation spectroscopy (XPCS) on a fragile binary model glass around and below $T_\text{g}$. We demonstrate a dynamically separated multi-stage relaxation behavior for aging times that cover 11 orders of magnitude and up to 10 microseconds in physical time, spanning from vibrational thermal dynamics at the shortest times, to the onset of an asymptotic limit of in simulations practically unreachable long time-scale decorrelation that is linked to the crystalline equilibrium phase formation (Fig.~\ref{fig:Figure1}a). In between, structural relaxation of the metallic model-glass is dominated by string excitations, whose mediated atomic displacements exhibit a clear magnitude gradient from frustrated amorphous domains to the continuously growing and eventually system-spanning interconnected network of low-energy motif clusters. Separating the decorrelation contributions by structural identification uncovers a traceable structure-dynamics correspondence that is defined by the microstructure of the glass.

\section*{Multi-stage structural decorrelations}

\begin{figure*}[!ht]
\centering
\includegraphics[width=0.8\textwidth]{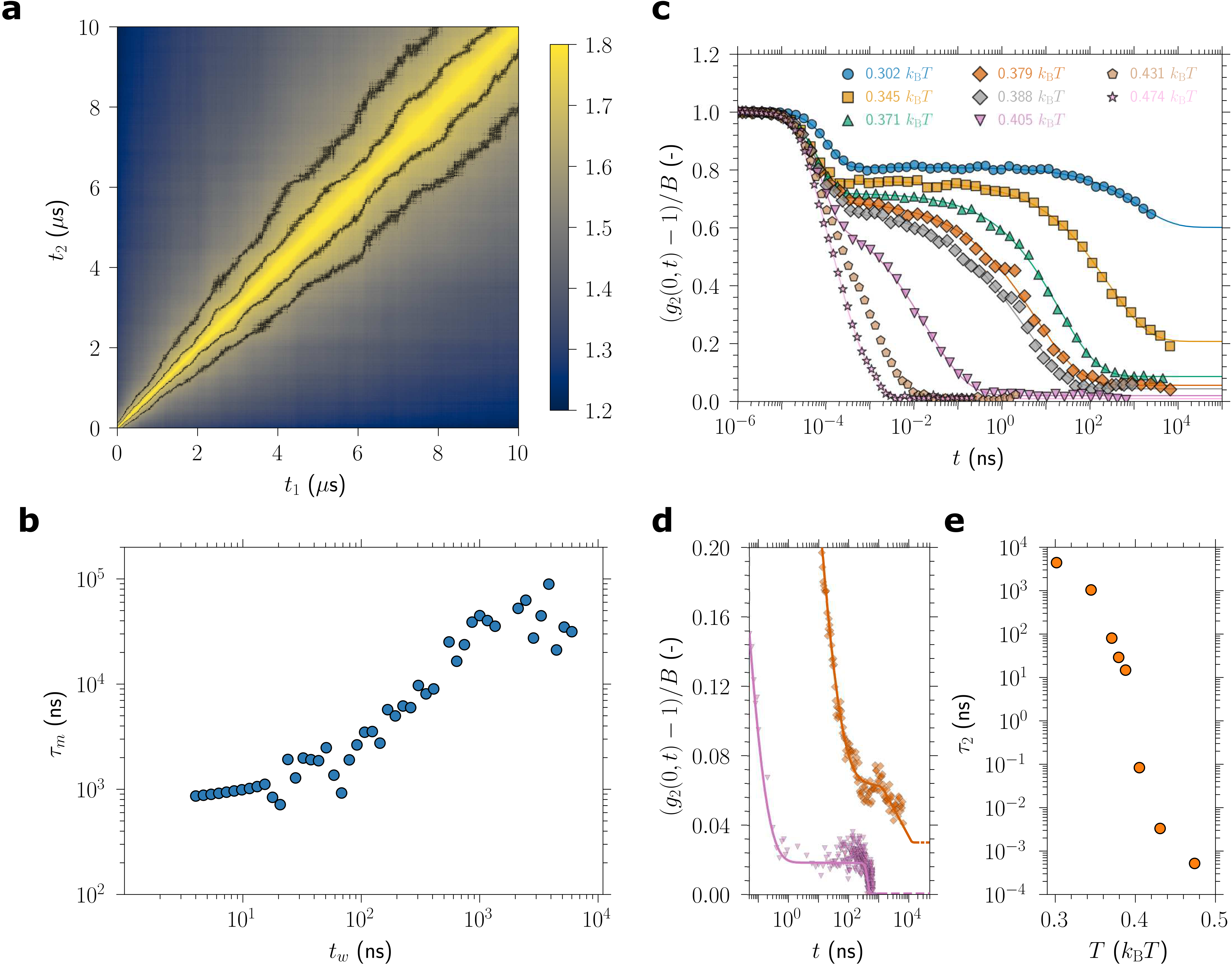}
\caption{(a) Exemplary two-times correlation function, $C(t_1, t_2)$, of the simulated binary model glass at a temperature of 0.345 $k_\text{B}T$, which is around 0.91 $T_{\text{g}}$. 
The two-time correlation function is calculated from the 3D diffraction patterns of the atomic structure of the model glass, which computationally mimics the x-ray photon correlation spectroscopy. The iso-contrast lines are plotted for values of $C(q, t_1, t_2) = 1.7$ and $C(q, t_1, t_2) = 1.6$, respectively. The color code indicates the values of the TTCF $C(q, t_1, t_2)$. (b) Momentary decorrelation time $\tau_m$ as a function of waiting time $t_w$. (c) The one time correlation function $g_2(0, t)$ at different temperatures around $T_{\text{g}}$.  (d) a zoomed-in onto the one-time correlation function at $T = 0.379 k_\text{B}T$ and $T = 0.405 k_\text{B}T$. The solid lines are fits to the sum of two KWW equations. (e) The decorrelation time $\tau_2$ as a function of temperature. 
}
\label{fig:Figure2}
\end{figure*}

Figure~\ref{fig:Figure2}a shows a two-time correlation-function (TTCF, $C$) constructed from simulated diffraction patterns (covering $7 < q < 8$ $\sigma^{-1}$) along a 10~\text{$\mu$}s long isotherm at a temperature equivalent to 0.91 $T_\text{g}$ ($\approx$ 0.345 $k_\text{B}T$) of the model glass ($T_\text{g}$ = 0.379 $k_\text{B}T$). For details on the TTCF, see Methods section. Iso-contrast lines for $C(q, t_1, t_2) = 1.7$ and $C(q, t_1, t_2) = 1.6$ are superimposed with the TTCF, highlighting heterogeneous or intermittent aging dynamics~\cite{Ruta2012PRL, Wang2024Acta} in the glassy solid domain, in which, contrary to stationary dynamics, the correlation varies over short time-windows with waiting time $t_w$. Figure~\ref{fig:Figure2}b and Fig.~S3 demonstrate this quantitatively via the momentary decorrelation time $\tau_m(t_w)$, indicating, despite noticeable fluctuations, an expected overall slowing down of the dynamics during energy minimization and the therewith increasing fraction of stable structural motifs~\cite{Wang2024Acta}.
Additional TTCFs with a qualitatively similar behavior are shown in Fig.~S1 for additional temperatures.

Alternatively to a time-window average producing $\tau_m(t_w)$, we calculate the normalized one-time correlation function $g_2(t_w, \Delta t) = \frac{\langle I(t_w)I(t_w + \Delta t) \rangle}{\langle I(t_w) \rangle^2}$, which is based on a delay time $\Delta t = 2$~ns and multiples thereof between the diffraction intensity in three-dimensional $q$-space (Fig.~\ref{fig:Figure1}b).
In other words, this represents an intensity cross-correlation of speckle intensity of the initial starting structure of the model glass from $t=0$ and forward in time. For the obtained temperature-dependent decorrelations, the resulting $g_2(t = 0, t)$ were fitted to the sum of two KWW-functions. The detailed fitting procedure is described in the Methods section, and resulting decorrelation times, $\tau_1$ (first decorrelation step), $\tau_2$ (second decorrelation step), their stretching exponents $\beta_1$ and $\beta_2$ are summarized in Fig.~\ref{fig:Figure2}e  and Fig.~S2. Clearly, shorter decorrelation times characterize higher temperatures, which is not only the case for $\tau_1$ related to thermal vibrations (Fig.~S2), but more importantly also for the second decorrelation due to aging-induced structural evolution. The latter is mediated by string-like collective excitations (Fig.~\ref{fig:Figure1}a) that have been revealed both in the undercooled liquid~\cite{Donati1998PRL} and the glassy solid domain~\cite{Derlet2021ActaMat, Gao2025NatP}. 
The normalized string activity exhibits a broad maximum whose position shifts systematically with temperature, indicating that string-mediated transport occurs over an extended temporal window rather than at a sharply defined characteristic time. The regime of enhanced string activity consistently overlaps with the second decorrelation stage of $g_2(t)$, providing direct support for the assignment of $\tau_2$ to dominant string-mediated glassy network evolution (see Fig.~S4).

Whilst not immediately apparent from the TTCF (Fig.~\ref{fig:Figure2}a), the remarkable feature of $g_2(t)$ is the absence of full decorrelations at long time scales. Indeed, leveraging simulations spanning 10~\text{$\mu$}s of physical time, a third decorrelation plateau $\neq 0$ becomes asymptotically visible even at the lowest temperatures. Being most pronounced in the glassy solid domain for T~$<$~0.379~$k_\text{B}T$, the data shows that the underlying cooperative-collective transport, typically contextualized as $\beta$-relaxations~\cite{Gao2025NatP, Riechers2024}, becomes sufficiently quenched out to be beyond practically achievable simulation times. Therefore, a decorrelation to the third plateau can only be unambiguously captured at and above $T_g$, as demonstrated for $T = 0.379~k_BT$ and $T = 0.405~k_BT$ (Fig.~\ref{fig:Figure2}d), where kinetic freezing and breaking of ergodicity~\cite{Berthier2023NP} in the studied model glass is still associated with accessible relaxation events within the simulated time window.

Given the incomplete decorrelation at the lowest temperature, and, as we will see later, the fact that fragments of the atomic-scale structure have not yet reached the expected equilibrium crystalline configuration, a third characteristic relaxation time $\tau_3$ must exist for the present system at t $>$ 1 $\mu$s. Being defined structurally at a later stage, the third decorrelation plateau provides strong evidence for multiple characteristic relaxation modes in even the simplest glass formers. This is very much compatible with the increasing number of secondary relaxation modes discovered for metallic glasses, or a quite general $n$-stage relaxation behavior in glassy solids~\cite{Yelash2012EPL, Solar2017JCP, Zhang2018PNAS}.

\section*{Spatially-correlated structural dynamics}

\begin{figure*}
\centering
\includegraphics[width=0.8\textwidth]{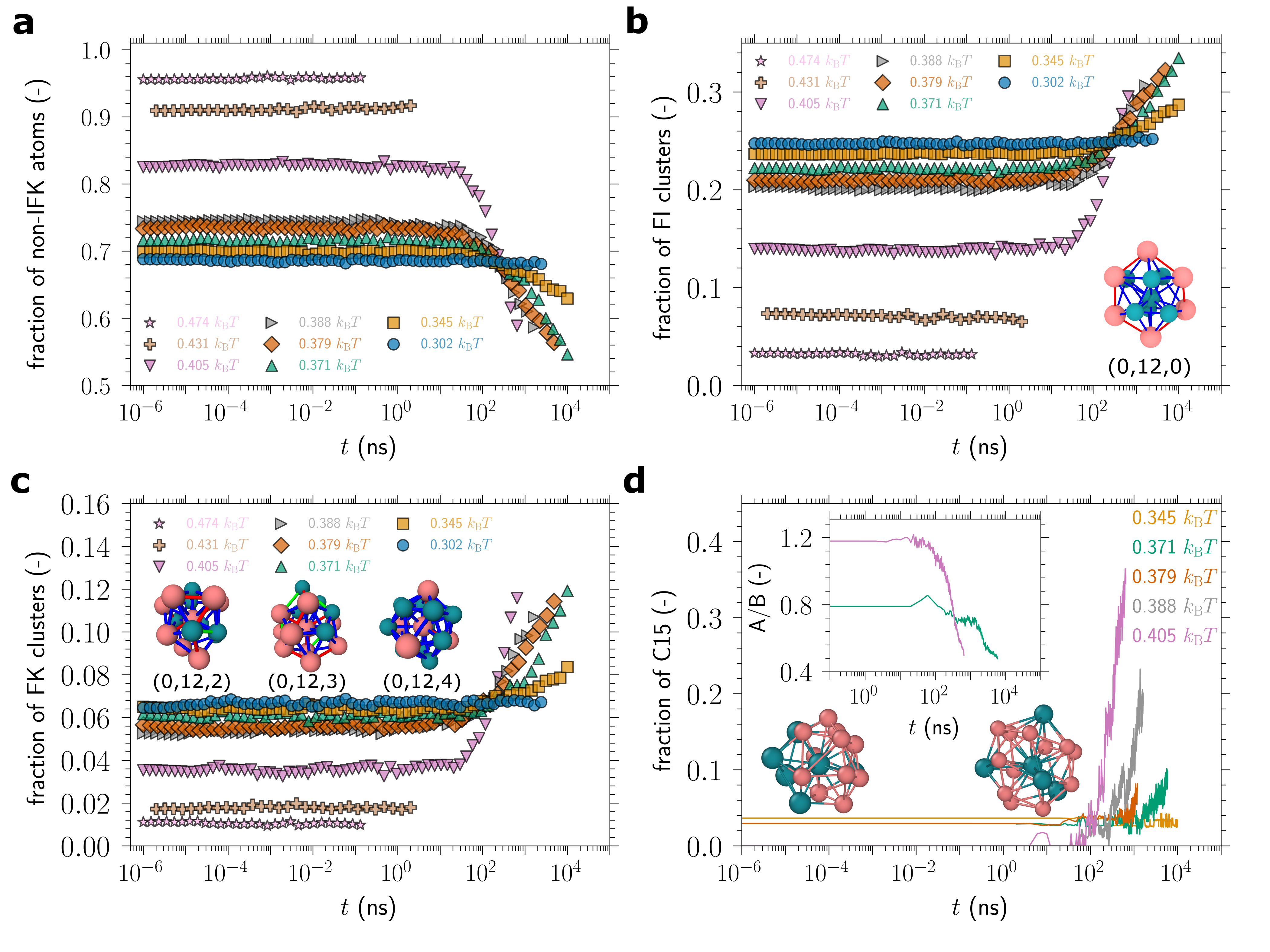}
\caption{(a) The time evolution of the fraction of the atoms that are not part of the IFK-network at different temperatures. (b) and (c) show the time evolution of the fraction of full icosahedra (0,12,0) and Frank-Kasper polyhedra (0,12,2), (0,12,3), and (0,12,4), respectively. The insets of plots (b) and (c) are representative snapshots of such motifs. (d) Fraction of perfect C15 Laves phases in the IFK-network at four selected temperatures. The inset in (d) shows the change in the ratio of A to B counts of the atoms identified as perfect C15 Laves phases at the final simulation step for two selected temperatures. A cutoff distance of 1.4~$\sigma$ was used to define neighboring atoms, which corresponds to the values of the first minimum of the radial distribution function.}
\label{fig:Figure3}
\end{figure*}

To shed light onto the uncovered multi-stage relaxation behavior of the model glass, the time- and temperature-dependent evolution of the structure is now evaluated. Figure~\ref{fig:Figure3} depicts the annealing-dependent fractional changes of the glass structure, differentiated by an amorphous component (Fig.~\ref{fig:Figure3}a), full icosahedral (0,12,0) motifs (FI, Fig.~\ref{fig:Figure3}b), and Frank-Kasper polyhedra (0,12,2), (0,12,3), (0,12,4) (FK, Fig.~\ref{fig:Figure3}c). The time evolution of fractions of each motif type is strongly temperature dependent, due to entropy domination at higher temperatures that gradually transitions to an enthalpy-dominated Gibbs free energy with decreasing temperature, during which stable low-energy motifs emerge in larger quantities. The slow increase in FK- and FI-motifs, occurring only after several hundred nanoseconds to a microsecond, despite a relatively high mean-squared displacement (MSD) ($\left< \Delta r^2(t) \right> \approx 4 \sigma^2$ at $t = 1~\mu s$ and $T = 0.345~k_\text{B}T$, Fig.~S8), indicates that the structural reorganization is not simply correlated with individual particle mobility. When comparing across the panels of Fig.~\ref{fig:Figure3} that cover 11 orders of magnitude in time, the IKF-motif growth emerges strongly at $\approx 100$~ns. However, closer inspection reveals clear temperature-dependent differences in onset times and rates of IKF-motif evolution. For example, FI fractions increase beyond 20 ns at $T = 0.405~k_\text{B}T$ but only after 70 ns at $T = 0.371~k_\text{B}T$. This structural evolution requires spatially cooperative transport and motif reorganization, which in fact drives an increasing structural connectivity and therefore the formation of a glassy microstructure. 

Asymptotically, the studied model glass is expected to crystallize into an AB$_2$ C15 Laves phase~\cite{Pedersen2007}, with A atoms on a cubic-diamond lattice occupying FK-sites and B atoms in corner-sharing tetrahedra occupying FI-sites. This is partly reflected in the change of their fractions, but it is important to understand that the emerging interconnected and system-spanning network of FI- and FK-motifs, the IFK-network (Fig.~\ref{fig:Figure1}c), is by definition not a crystalline Laves phase. Tracing structurally perfect C15 Laves motifs at selected temperatures demonstrates an increasing fraction of such ordered structures within the IFK-network beyond 100 ns, as seen in Fig.~\ref{fig:Figure3}d. This occurs when much of the IKF-network has formed. With C15 Laves motif fraction, we consider here its share of the IFK network and not the full simulation box. For example, at $T = 0.405~k_BT$ and $t = 300$~ns, a C15 fraction of $\sim$30\% within the IFK network corresponds to only $\sim$6\% of all atoms in the system, given that the IFK network itself comprises approximately 30\% of the total at that time. 

Above $T_\text{g}$, the increase in the structurally-identified C15 fraction occurs quickly in spurts, whereas the lower temperatures show a distinctly slower increase in C15 fraction. Deep within the glassy solid at $0.302~k_BT$, the structural network evolution proceeds by an increasing inter-connectivity of FK- and FI-motifs within the studied time window. Investigating the corresponding chemical A/B-ratio of the structurally-identified C15 motifs reveals a delayed chemical ordering towards the expected A/B-ratio of 0.5. This is demonstrated in the inset in Fig.~\ref{fig:Figure3}d, where, for example, the A/B-ratio at $0.405~k_BT$ first clearly decays from its initial value at 200 ns when its structurally-identified domains within the IFK-network have already exceeded a fraction of 10~\%. In other words, first a topological ordering out of the IFK-network into C15 fragments occurs, after which continued structural transport facilitates the chemical ordering into the AB$_2$ C15 phase. This underscores that chemical ordering of dispersed C15 fragments represents the latest stage of physical aging but has only sufficient intensity to drive the third decorrelation at the highest studied temperatures for the accessible time window.

\begin{figure*}
\centering
\includegraphics[width=0.8\textwidth]{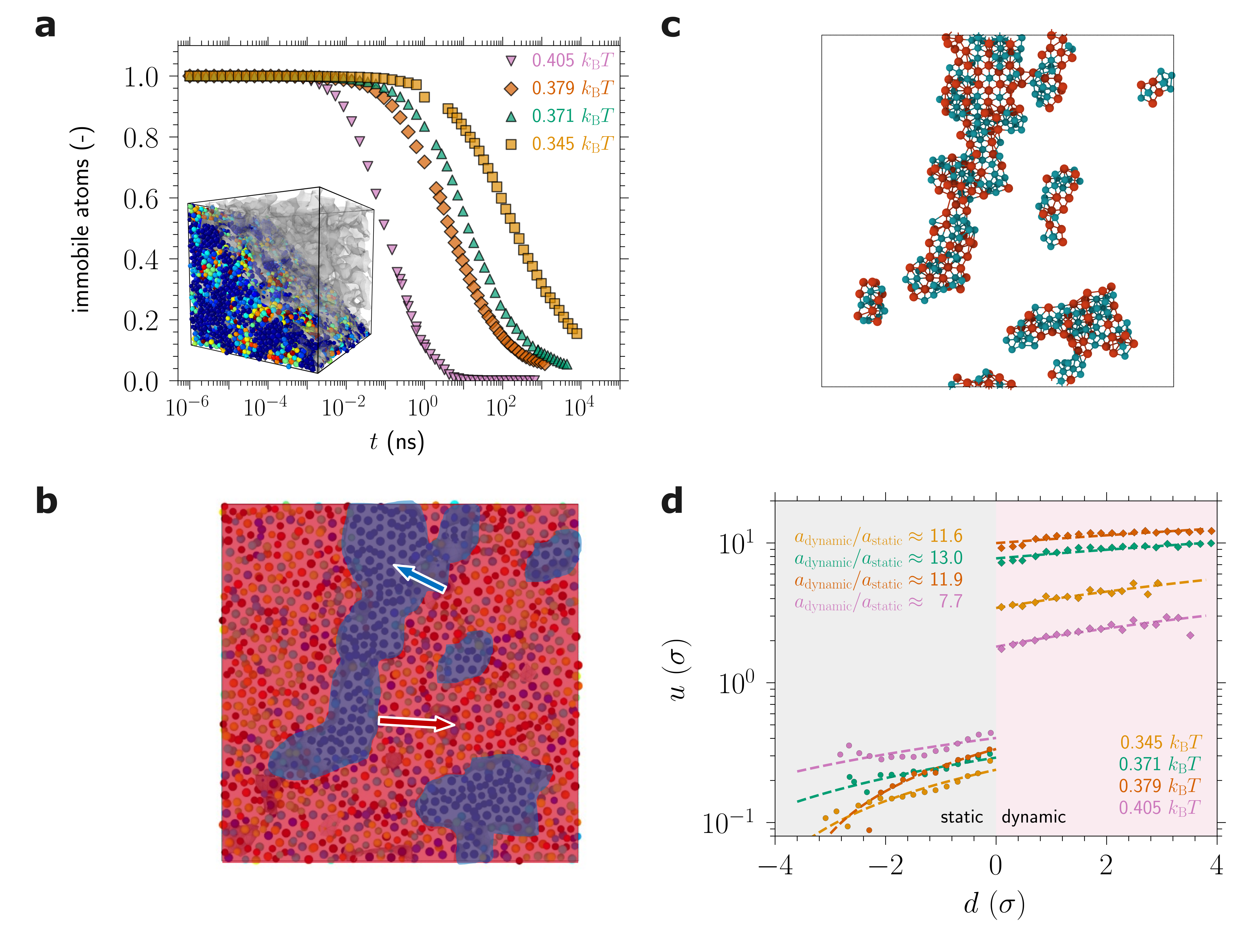}
\caption{(a) The time evolution of the fraction of atoms that are identified as immobile for four selected temperatures. The immobile atoms are labeled as atoms that moved a distance less than 0.6~$\sigma$ at time $t$. The inset in figure (a) shows the simulation box and a mesh that serves as a border between the clusters of immobile and mobile atoms. (b) Structural snapshot highlight a dynamical separation of the model glass. The dividing surface was constructed by separating atoms with displacement magnitudes below and above $0.6~\sigma$. This effectively binarized view is exaggerated by the set displacement threshold, but separating the less mobile component, show in (c) reveals a clear structural correspondence of with the between the IFK network. (d) Displacement of atoms at a distance $d$ away from the constructed dividing surface. The data was block-averaged using a bin size of 0.2 $\sigma$. Negative values of $d$ indicate that the distance is going inside the region of the immobile atoms, while positive values of $d$ are for the distances going inside the mobile region as highlighted with arrows in (b). The solid lines are fits to the unaveraged raw data (see Fig.~S7), and their slopes are indicative of a gradient in the displacement of the atoms. For the isotherms simulated at $T = 0.345, 0.371$, and $0.379~k_\text{B}T$ the snapshots where taken at $t =$ 1~$\mu$s and for $T = 0.405~k_\text{B}T$ the time was t = 0.4~ns.}
\label{fig:Figure4}
\end{figure*}

This evolving microstructure is expected to admit spatially-dependent transport properties that link dynamics to a three-dimensional structural network. By defining a displacement-magnitude cut-off based on Fig.~S6 of $d_{\text{c}} =$ 0.6~$\sigma$, mobile and immobile atoms are therefore tracked. The time-dependent evolution of the fraction of immobile atoms is summarized in Fig.~\ref{fig:Figure4}a for selected temperatures above, at, and below $T_\text{g}$. For $T\leq 0.379 ~k_BT$ even after several microseconds, a finite fraction of immobile atoms are present that exclusively belong to the IFK-motif population. This is associated with the presence of a more static low-energy motif network that is interspersed by dynamical channels in the relaxing glass. The inset in Fig.~\ref{fig:Figure4}a depicts this scenario by an example snapshot of the simulation box, which shows an atomic structure visualization with non-blue atoms being more mobile and an IFK-network representation.

The progressing formation of a strongly partitioned glass structure results in spatially heterogeneous relaxation behavior, in which a spatial correspondence is observed between low-displacement regions and the IFK motifs network, indicating that atoms in these motifs exhibit reduced mobility and leading to the time-scale separation of $g_2(t)$. It also contributes to quantifiable displacement gradients from the amorphous channels into the system-spanning low-energy motif network. Based on the above definition of mobile and immobile atoms, a dividing surface between dynamically active and non-active domains of the sample can be constructed. This is demonstrated in Fig.~\ref{fig:Figure4}b by a structural snapshot for $T = 0.345~k_\text{B}T$ ($4000~\text{K}$) and $t = 1~\mu$s. The essentially binarized appearance of Fig.~\ref{fig:Figure4}b is a result of block-averaging the unaveraged data of Fig.~S7 with a bin size of $0.2~\sigma$. Whilst exaggerated, it reveals the spatial discontinuity of the atomic displacement field, where the IFK network distinctly marks the, within the time-window of observation, immobile component (Fig.~\ref{fig:Figure4}c). This holds true for all temperatures, as revealed in Fig.~\ref{fig:Figure4}d by the atomic displacement, $u$, as a function of distance $d$ across the interface, where the sign indicates the direction. While the displacement away from the defined interface towards the center of the mobile region increases, it decreases towards the center of the immobile region. Quantified, the ratio of fitted positive-valued linear slopes $\frac{a_{dynamic}}{a_{static}}$ deviate markedly from unity as noted in Fig.~\ref{fig:Figure4}d for various temperatures. 

Even though dynamical gradients for metallic glasses have been revealed with electron-correlation microscopy near free surfaces~\cite{Zhang2018NatCom}, the here conducted analysis establishes a one-to-one mapping between distinct microstructural components and dynamical gradients with the bulk of the material. The evolution of the microstructure and the entailed network formation leads to a structurally and dynamically distinctly heterogeneous picture, where the string-mediated relaxation mechanism is the same in both network and non-network part. As will be seen in the following, it is this microstructure-dynamics coupling that underlies the multi-stage structural decorrelation of the model glass. 

\section*{Atomic-scale dynamics in different structural components}

\begin{figure}
\centering
\includegraphics[width=\columnwidth]{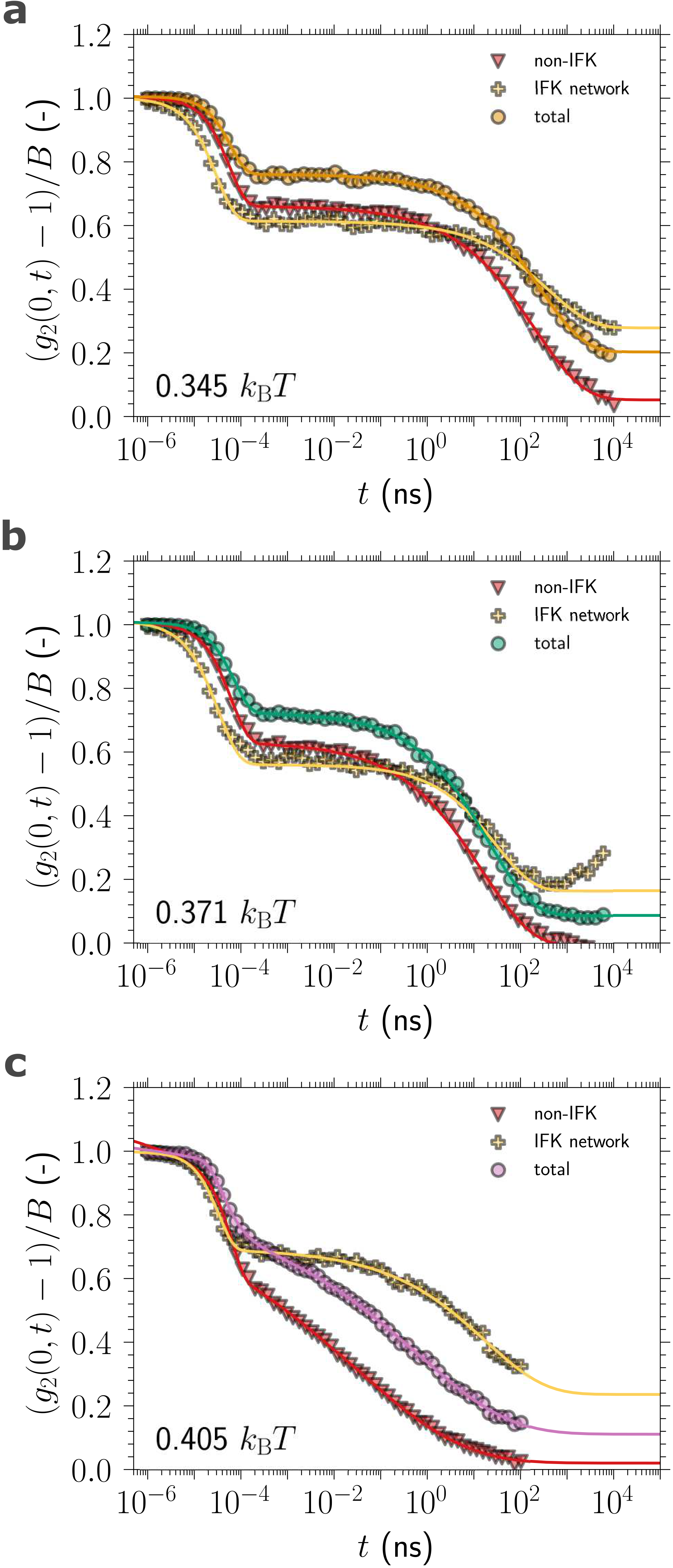}
\caption{Normalized $g_2(0, t)$ of the total sample involving all atoms, and its deconvolution to the contribution of amorphous and non-amorphous atoms at different selected temperatures. Solid lines through the data are best fits to the double KWW equations. At $T = 0.371~ k_\text{B}T$, $g_2(0, t>1\mu s)$ shows a recorrelation of the non-amorphous atoms. For the data in (c) at $T = 0.405~k_\text{B}T$, the reference to calculate $g_2(t)$ was selected as $t = 300$~ns, has a 30\% fraction of IFK motifs.}
\label{fig:Figure5}
\end{figure}

The dynamically well distinguishable components and the displacement gradient from mobile to static atomic environments now give insights into the multi-stage decorrelations of Fig.~\ref{fig:Figure2}c. To this end, $g_2(0,t)$ for the IFK and non-IFK regions was computed separately at 0.345~$k_\text{B}T$ and 0.371~$k_\text{B}T$ (Fig.~\ref{fig:Figure5}a and ~\ref{fig:Figure5}b). Remarkably, both structural components decorrelate distinctly different. At short times, the IFK motif-network decorrelation is faster than for the non-IFK structures, whereas this is opposite at longer times, when the decorrelation from the second to the third plateau occurs. This crossover in the decorrelation behavior between the two structural components is due to the rapid structural adjustment of the IFK-network: once minimized energetically via short-timescale relaxation, this stable network becomes structurally increasingly long-lived, leading to a relatively slower decorrelation in the microsecond regime of the amorphous domains. From the data in Fig.~\ref{fig:Figure3} and Fig.~\ref{fig:Figure5}, it becomes clear that the emerging third decorrelation plateau originates from the IKF motif-network gradually forming first topologically ordered and subsequently also chemically ordered crystalline C15 Laves structures. The insets in Fig.~\ref{fig:Figure3}d visualized this with a representative chemically disordered C15 motifs (left) that subsequently attains appropriate chemistry (right). Figure~S5 presents the temperature-dependent waiting-time statistics of hopping events for atoms belonging to the final Laves phase, indicating power-law scaling and therefore some correlated transport. This incipient crystallization is expected to contribute with an additional decorrelation characterized by $\tau_{3}$, the onset of which is in fact observed at the longest simulation times for $T = 0.405~k_\text{B}T$ and $T = 0.379~ k_\text{B}T$ and depicted in Fig.~\ref{fig:Figure2}d.

These results provide evidence of a marked timescale separation due to temperature-driven structural partitioning and, therefore, the formation of a glassy microstructure. This can be further substantiated by a structure-specific $g_2(0,t)$ for the model glass, now demonstrated for 0.405~$k_\text{B}T$ and intensity correlation to the structural state at $t = 300$~ns (Fig.~\ref{fig:Figure5}c). Effectively, this represents a different initial reference structure, similarly as using orders of magnitude different initial quench rates prior to the isotherm. The waiting time of $t = 300$~ns was chosen because at that time the IFK-network fraction is similar to that of all other lower temperatures (cf. crossing of datasets in Fig.~\ref{fig:Figure3}a-c). Now, a pronounced separation of time-scales becomes evident, such that even the KWW functional form no longer accurately describes the temporal decay correlations. At this point, the percolative and system-spanning IFK network unambiguously defines the long-time decorrelation plateau. Its own relaxation time, $\tau_3$, being beyond reach for state-of-the-art MD simulations, is expected to lead to further decorrelation. Asymptotically, this must be accompanied by extensive Laves-phase crystallization. 

\section*{Conclusions and Outlook}

The here studied binary model glass is simple, yet, if sufficiently long studied, thermally-activated structural relaxation drives the formation of an extended microstructure with clearly defined domains. Coupled to this structural partitioning are in-time separated decorrelation domains, each characterized by a dominant mechanism and time scale. These are $\tau_1$ (thermal vibration), $\tau_2$ (strings), and $\tau_3$ (topological/chemical ordering), the latter of which is only accessible in part beyond 200 ns for 0.405~$k_\text{B}T$ and beyond 1 $\mu$s for 0.379~$k_\text{B}T$ but are clearly linked to the ordering of the IFK-network. Whilst one may conclude that this final decorrelation due to crystallization is the asymptotic limit of the relaxation path of the studied model glass, it cannot be excluded that the formed incipient crystalline nuclei are sub-critical and may potentially dissolve again at inaccessible times. Alternatively, full crystallization of the sample into AB$_2$ C15 Laves phase may occur, while the remaining 25\% A atoms could crystallize into a closed-packed structure. This would indeed entail another decorrelation of $g_2(t = 0, t)$, but due to the roughness of the C15 interface further ordering of an amorphous A-rich region has been shown to be frustrated~\cite{Derlet2020JALCOM}. Consequently, residual amorphous domains should persist asymptotically, depending on the interfacial energy landscape and kinetics. The key finding is thus not the first signatures of incipient crystallization across all temperatures in both the undercooled liquid regime and the glassy solid domain, but the uncovering of distinct, dynamically separated decorrelation regimes with specific dominant structural mechanisms, including a pre-crystallization stage of topological and chemical ordering that produces a well-separated plateau in $g_2(t)$.   

To assess whether the identified multistage relaxation behavior is specific to the binary Lennard-Jones model, additional simulations of a ternary Zr$_{46}$Cu$_{46}$Al$_8$ metallic glass were conducted using an embedded atom model (EAM) potential~\cite{Cheng2009} - an alloy composition that is well studied experimentally. At $T = 850$~K ($\approx 1.13~T_g$), $g_2(t)$ exhibits a qualitatively similar multistep decorrelation, reaching an intermediate plateau at approximately 0.2 (Fig.~S10). This, together with the robust multistep decorrelation of the binary model irrespective of initial IFK-network distribution or fraction at $t_0$, demonstrates that the uncovered hierarchical relaxation trajectory is neither linked to specifics of the binary model nor the glass preparation, but reflects a more general physics of metallic (model) glass aging.

To better reconcile the here found multi-stage structural decorrelation in a fragile binary model-glass, the findings need to be contextualized in an experimental setting. Despite the vastly different time scales, the isothermal simulations and probed structural dynamics represent well a frequency-dependent mechanical spectroscopy experiment at constant temperature. Maxima in the loss modulus would thus be expected at frequencies corresponding to the plateau-specific relaxation times. At a laboratory time scale, this is nothing else than the emergence of multi-peak loss spectra obtained for metallic-glass alloys, if probed up to sufficiently high frequency for long enough time. Besides some binaries, such as CuZr, NiP, or NiNb, metallic glasses are mostly ternary, quaternary, or quinary systems, whose atomic-scale motif clustering and local chemical ordering continue to remain unknown. However, qualitatively complex clustering and network evolution similar to the model glass must accompany enthalpy minimization during relaxation, as now becomes increasingly evident from 4D transmission electron microscopy~\cite{Ortiz2025,Riechers2026}. Given the limited high-frequency domain of mechanical spectroscopy data of metallic glasses, the here conducted simulations therefore suggest the existence of additional yet uncovered fast relaxation modes in the ms-regime and below.

In view of the now routinely exploited XPCS method to quantify temperature- and stress-driven structural dynamics of metallic glasses, it is indeed remarkable that the entire literature body reports, with one exception~\cite{Frey2025NatCom}, a single decorrelation. The corresponding relaxation times range from seconds in the undercooled liquid domain to $10^4$~s in the glassy solid domain, without any reported evidence of multiple isothermal decorrelations. Based on the data in Fig.~\ref{fig:Figure2}c, it becomes clear that especially for well-relaxed metallic glasses, many orders of time-scale are required to capture sequential transport domains and their decorrelations. Figure~S9 supports this conclusion via two-stage decorrelations of $g_2(t = 0, t)$ from our exceptionally long XPCS experiments, one of which was recently reported in Ref.~\cite{Riechers2024} but with considering $g_2(t=t_1-t_2,t_2)$ instead of $g_2(t=0,t)$. This suggests that most experimentally probed metallic-glass structures are always in a microstructural state that, at the given speckle-pattern resolution and for the duration of the experiments, admit atomic-scale rearrangements captured collectively in one effective decorrelation time scale regime.

Based on our findings, we thus anticipate that if probed with sufficiently high time resolution over long enough times, a multi-stage decay of the ISF for highly unrelaxed metallic glasses must exist, where dominant and characteristic structural mechanisms identify the various decorrelations. This is now experimentally in reach at the worldwide emerging upgraded coherent scattering beamlines at fourth-generation synchrotrons. Alternatively, annealing-induced glass-glass separations should promote multi-stage one-time correlation functions in XPCS experiments due to an expected dynamical separation in the polymorphic glass~\cite{Shen2022ActaMat}. 

\section*{Method}
\subsection*{Glass model and isothermal annealing}
We perform molecular dynamics (MD) simulations of a model binary glass former system. The sample is made of A and B atoms with equal fractions (50:50). The atoms are interacting with a Lennard-Jones potential, eq.~\eqref{eq:lj}, with the parametrization from Wahnström~\cite{Wahnstrm1991}. This model enables the study of glasses and glass-forming liquids without any chemical bias, where only volume frustration due to size disparity is present.

The two atoms A and B are separated by a distance $r$ and their interaction energy $V_{ij}(r)$ is given by,

\begin{equation}
\label{eq:lj}
V_{ij}(r) = 4\varepsilon\left(\left(\frac{\sigma_{ij}}{r}\right)^{12} - \left(\frac{\sigma_{ij}}{r}\right)^{6}\right), 
\end{equation}

where $i$ and $j$ are two different atoms. The potential parameters are as follows: $\varepsilon$ is set to 1 and used to define the energy unit, $\sigma_{\text{AA}} = 1$ is used for the unit of length, $\sigma_{\text{BB}} = \frac{5}{6}~\sigma_{\text{AA}}$ and $\sigma_{\text{AB}} = \sigma_{\text{BA}} = \frac{11}{12}~\sigma_{\text{AA}}$. The atomic masses of the two atoms are selected as $m_1/m_2 = 2$. The potential is truncated and shifted to zero at 2.5$~\sigma$. When assuming a realistic metallic system, such as a prototypical Zr-Cu alloy, one simulation step is approximately 0.2 fs for $T \ge 0.405~k_\text{B}T$ (4700~K) and 0.4 fs for $T < T = 0.405~k_\text{B}T$. This timescale described above is used to present the time-dependent data.

The samples were prepared by placing 32000 atoms in a cubic periodic box and heating it to generate an equilibrium liquid state. The equilibrated liquid is then quenched using a rate of 5 $\times$ 10$^9$~K/s to the temperatures of interest above and below $T_\text{g}$. Temperatures are expressed in units of $k_\text{B}T$ and their corresponding values in Kelvin are given between parentheses in the following: $T$ = 0.302 (3500~K), 0.345 (4000~K), 0.371 (4300~K), 0.379 (4400~K), 0.405 (4700~K), 0.431 (5000~K), 0.474 (5500~K). Isothermal annealing up to an equivalent of 10 $\mu$s of physical time are considered. The analysis of the structure was performed on atomic configurations recorded logarithmically for times shorter than 1 ns and at 2 ns intervals for times longer than 1 ns.

To validate the generality of the conclusions beyond the model binary system, we additionally perform MD simulations of a more realistic ternary metallic glass (Zr$_{46}$Cu$_{46}$Al$_8$, at.\%) using an embedded-atom potential parameterized by Cheng et al.~\cite{Cheng2009}. A cubic periodic box was filled with 64000 atoms, heated to 2000~K to generate an equilibrium liquid, then quenched at $10^{9}$~K/s to about $\sim 1.13~T_\text{g}$ ($T =$ 850~K, with $T_\text{g} \approx$ 750~K) for isothermal annealing up to 2.5$\mu$s. Time steps of 0.1 fs, 1 fs, and 2 fs were used for $t <$ 1 ps, $t <$ 1 ns, and $t >$ 1 ns, respectively.

All simulations are performed at a constant pressure of 0 MPa. The temperature and pressure control is handled by using a Nose-Hoover thermostat and barostat~\cite{Shinoda2004PRB}. All MD simulations are performed using LAMMPS~\cite{Thompson2022}, the analysis using custom Python scripts, and all atomic visualizations are generated with OVITO~\cite{Stukowski2009}.

\subsection*{Computational x-ray photon correlation spectroscopy}
The diffraction patterns were calculated from the atomic positions using eq.~\eqref{eq:Sq}

\begin{equation}
\label{eq:Sq}
I(\bm{q},t) = \frac{1}{N}\left|\sum_i^N e^{-i\bm{q}\cdot\bm{R}_i(t)}\right|^2,
\end{equation}
with \bm{$q$} being the reciprocal space wave vector, $N$ the total number of atoms, and \bm{$R_i$} is the position vector of an atom $i$ in real space. In the specific case of our simulations, an equal weight was given for all atoms by setting the form factor to 1. 
Due to the finite size of the system in the simulations and the need to minimize artifacts that can appear due to the periodic boundary conditions, the choice of the $\bm{q}$-vectors were restricted to $\frac{2\pi}{L}[n_x, n_y, n_z]$, with $L$ is the length of the simulation box, and $n_x$, $n_y$, and $n_z$ are integers. The simulated box size is $30\times30\times30~\sigma^3$, which gives an approximate $q$-space resolution of around $0.2\sigma^{-1}$/pixel. This, when compared to typical experimental detector resolution of around 10$^{-4}$ \text{\AA}$^{-1}$, reflects the small scattering volume and resolution, which is offset by analyzing a much larger $q$-range for speckle pattner cross correlations (see below). 

Although the intermediate scattering function (ISF) can in be calculated over the same time window, the simulated XPCS approach is preferred for two reasons. First, it can be directly compared to experimental results, because it produces similar observables. Following this approach, there is also no need to rely on the applicability of the Siegert relation, which assumes Gaussian particle dynamics, which might not hold for the out-of-equilibrium aging of metallic glasses, where the dynamics are non-Gaussian and non-ergodic. Second, $g_2(t)$ retains sensitivity to dynamic heterogeneity and intermittency, a sensitivity that is attenuated in ensemble-averaged ISFs.

The 3D diffraction spheres obtained from the atomic trajectories are then used to compute an intensity-intensity cross correlation to get a two-time correlation function (TTCF), given by eq.~\eqref{eq:TTCF} 

\begin{equation}
\label{eq:TTCF}
C(\bm{q}, t_1, t_2) = \frac{\left<I(\bm{q}, t_1)I(\bm{q}, t_2)\right>_q}{\left<I(\bm{q},t_1)\right>_q\left<I(\bm{q}, t_2)\right>_q}
\end{equation}
where $\left<\dots\right>_q$ denotes an average over all selected $q$-pixels, with $q$ ranging between 7 and 8 $\sigma^{-1}$, which corresponds to a $q$ range between 2.35 and 2.7 \text{\AA$^{-1}$}, when translating the units into a realistic Cu-Zr metallic glass sample.
From the obtained TTCFs, one-time correlation functions, $g_2(t = 0,t)$, are obtained by setting the time $t$ and moving horizontally starting from the diagonal at that time. Unless stated otherwise, the one-time correlation function $g_2(t_0, t)$ is calculated with a fixed reference time $t_0 = 0$~ns, which corresponds to the start of the isothermal anneal immediately after the quench. (e.g., Fig.~5c uses $t_0 = 300~\text{ns}$ at 0.405~$k_BT$). The only exception to this approach is given by the data presented as $\tau_m$ in Fig.~\ref{fig:Figure2}b with varying reference times $t_0$ as described below, and in Fig.~\ref{fig:Figure5}c, which has the reference time $t_0 = 300~\text{ns}$ at $T=0.405~k_BT$. Each curve plotted in this paper represents a single annealing trajectory of a simulation box containing 32000 atoms. Such trajectories are overall reproducible but differ in fine details of energy minimization.

\subsection*{Fitting and extraction of relaxation times}
The momentary decorrelation time $\tau_m(t_w)$ is extracted from the TTCF by fitting a KWW function to a time window of fixed width centered on the waiting time $t_w$, and represents the instantaneous structural relaxation timescale at that point during the isotherm.

The $g_2(0, t)$ data was fitted to the sum of two Kohlrausch–Williams–Watts (KWW) (eq.~\eqref{eq:KWW}) functions to account for the multiple steps of the relaxation

\begin{equation}
\label{eq:KWW}
g_2(0, t) - 1  = Ae^{-2(t / \tau_1)^{\beta_1}} + Be^{-2(t / \tau_2)^{\beta_2}} + C,
\end{equation}

where $\tau_i$ and $\beta_i$ are the characteristic relaxation time and the stretching exponent, respectively. $A$ and $B$ are coefficients that determine the amplitude of the exponential decay components in the correlation function, representing the contribution of each decay process. $C$ is the baseline that accounts for the background level. 

Intensity autocorrelation functions computed separately for A and B atom types yield nearly identical decorrelation traces, indicating that the overall relaxation dynamics are governed primarily by local topology and motif environment rather than atomic mass or species identity.

\subsection*{Voronoi analysis and Laves phase identification}

A Voronoi analysis is performed according to the method proposed by Derlet~\cite{Derlet2020PRM}, which allows for the calculation of Voronoi motifs while removing nonphysical bonds found in the sample and taking into account the atomic size as well, which is suitable for poly-disperse glasses. The results of the Voronoi analysis classify the structure into different polyhedral motifs, characterized by the number of faces $n$ and $k$ edges in these motifs. This lead to labeling ($n_{k=4}$, $n_{k=5}$, $n_{k=6}$). Following such notation, a full icosahedron is (0,12,0), and the Frank-Kasper polyhedra are (0,12,2), (0,12,3), and (0,12,4). The here studied model glass crystallizes in a C15 Laves phase, which is a cubic phase in the form AB\textsubscript{2}. Here, A atoms occupy FK-sites on a cubic diamond lattice, and B atoms occupy FI-sites in corner-sharing tetrahedra. A custom script was written to analyze the existence of AB$_2$ C15 Laves structure within the samples. The number of neighbors of each atom was computed using a fixed radial cutoff $r_\text{cut} = 1.45~\sigma$. Subsequently, atoms are evaluated according to their type and local environment that match the C15 Laves phase described above. The detected Laves structures were further confirmed using the tool developed by Xie et al.~\cite{Xie2021}, which works in principle similarly to the common neighbors analysis to identify Laves phases, primarily C14 and C15. This analysis method identifies motifs with the structure of the Laves phase, including both stoichiometric and off-stoichiometric C15 structures. 

%Topologically and chemically ordered C15 Laves phases were further detected by applying additional constraints on the structural motif search: each A atom must have 4 A-neighbors and 12 B-neighbors, and the angles between A atoms must approximate the tetrahedral angle, allowing a 10\% deviation to account for thermal distortions.

\subsection*{Atomic mobility and displacement gradients}

At a given time, the displacement of the atoms can be calculated with respect to a known reference configuration (e.g., $t$ = 0). Such analysis allows for the classification of atoms into mobile and immobile atoms by defining a cutoff in the displacement distribution of all atoms. This cutoff was chosen as 0.6~$\sigma$, which corresponds to a minimum in the displacement distribution after 1 $\mu s$ of simulations (See Fig.~S6). Once these atoms are classified, the immobile atoms (which are found as interconnected clusters) are used to create a mesh that separates the mobile region from the immobile one. As a next step, the distance of atoms from this mesh is calculated and used to plot the displacement as a function of the distance towards the immobile or mobile regions. This reveals if a gradient in the displacement exists or not.

\subsection*{String-like displacement}
Correlated string-like displacements were identified using the method of Ref.~\cite{Donati1998}, according to which a displacement is classified as a string segment when the position of atom $i$ at time $t$ lies within a tolerance $\Delta r = 0.12\sigma$ of the initial position of a different atom $j$, i.e., $\left|\mathbf{r}_i(t) - \mathbf{r}_j(0)\right| \leq \Delta r$. The reference configuration is defined at $t=0$, and the analysis is performed for a fixed lag time $t$. Candidate pairs are restricted to atoms separated by less than $1.45\sigma$ in the reference configuration, corresponding to the first minimum of the radial distribution function; this criterion is applied uniformly to all pair types. String segments are constructed by connecting compatible pairs such that the final position of one atom matches the initial position of the next, forming non-branching chains with each atom assigned to at most one chain. The primary observable is the chain length, defined as the number of atoms in each string.

\section*{Acknowledgements}
The authors gratefully acknowledge Peter M. Derlet for supporting the technical implementation of the simulations, for fruitful discussions, and for interpretation of the obtained data. R.M. thanks BAM for institutional and financial support to conduct this research.

\end{document}